\documentclass[runningheads]{llncs}
\usepackage[T1]{fontenc}
\usepackage{graphicx}
\usepackage{hyperref}       
\usepackage{url}            
\usepackage{booktabs}       
\usepackage{amsfonts}       
\usepackage{nicefrac}       
\usepackage{microtype}      

\usepackage{doi}
\usepackage{amssymb}
\usepackage{multirow}
\usepackage{tabularx}
\PassOptionsToPackage{hyphens}{url}
\usepackage{float}
\usepackage{textcomp}
\usepackage{units}

\usepackage[backend=biber,style=numeric]{biblatex}
\usepackage{enumerate} 
\usepackage{wrapfig}
\usepackage{caption}
\usepackage{multicol}
\usepackage{hyperref}
\usepackage{gensymb}
\usepackage{eurosym}
\usepackage{amsmath}
\usepackage{lscape}
\usepackage{bm}
\usepackage{xcolor}
\begin{document}
\title{Multi-task Learning Approach for Intracranial Hemorrhage Prognosis}
%
%
\newcommand*\samethanks[1][\value{footnote}]{\footnotemark[#1]}
\author{Miriam Cobo\inst{1,2,3}\orcidID{0000-0001-9208-9062} \and
Amaia Pérez del Barrio\inst{4}\orcidID{0000-0003-1613-6159} \and
Pablo Menéndez Fernández-Miranda\inst{5,6}\orcidID{0000-0003-4742-8955}
\and
Pablo Sanz Bellón\inst{7}\orcidID{0000-0001-6237-0611}
\and
Lara Lloret Iglesias \inst{1}\thanks{Both authors share Senior authorship.}\orcidID{0000-0002-0157-4765}
\and
Wilson Silva\inst{2,3}\samethanks\orcidID{0000-0002-4080-9328}}
\authorrunning{Cobo et al.}
%
\institute{Advanced Computing and e-Science Group, Institute of Physics of Cantabria (IFCA), CSIC - UC, Santander, Spain,
\email{cobocano@ifca.unican.es}\\ 
\and AI Technology for Life, Department of Computing and Information Sciences, Utrecht University, Utrecht, Netherlands \and Department of Radiology, The Netherlands Cancer Institute, Amsterdam, Netherlands \and 
Servicio de Radiología, Hospital Reina Sofía, Tudela, Navarra, Spain \and
Servicio de Radiología, Hospital Universitario Rey Juan Carlos, Móstoles, Spain \and 
Departamento de Tecnologías de La Información, Universidad CEU San Pablo, Madrid, Spain \and
Servicio de Radiología, Hospital Universitario Marqués de Valdecilla, Santander, Spain}
\maketitle              
\begin{abstract}

Prognosis after intracranial hemorrhage (ICH) is influenced by a complex interplay between imaging and tabular data.
Rapid and reliable prognosis are crucial for effective patient stratification and informed treatment decision-making. In this study, we aim to enhance image-based prognosis by learning a robust feature representation shared between prognosis and the clinical and demographic variables most highly correlated with it. Our approach mimics clinical decision-making by reinforcing the model to learn valuable prognostic data embedded in the image. 
We propose a 3D multi-task image model to predict prognosis, Glasgow Coma Scale and age, improving accuracy and interpretability. Our method outperforms current state-of-the-art baseline image models, and demonstrates superior performance in ICH prognosis compared to four board-certified neuroradiologists using only CT scans as input. We further validate our model with interpretability saliency maps.
Code is available at \textit{https://github.com/MiriamCobo/MultitaskLearning\_ICH\_Prognosis.git}.

\end{abstract}
\section{Introduction}

Intracranial hemorrhage (ICH) is a leading cause of death and disability worldwide, characterized by high mortality rates and significant long-term neurological impairments \cite{witsch2021prognostication}. This condition poses a substantial burden on healthcare systems, and presents complex challenges in terms of timely diagnosis, effective treatment, and rehabilitation strategies \cite{gregorio2019assessment}. The incidence of ICH is projected to rise due to aging populations and increasing prevalence of risk factors such as hypertension, coagulopathy, and cerebral amyloid angiopathy \cite{magid2022cerebral}. 
Yet, prognostic predictors of ICH are significantly under-explored, hindering patient stratification by severity and the evaluation of the efficacy of emerging therapeutic interventions \cite{predispose2018unmet}. This prognostic uncertainty sometimes leads to a lack of consensus among clinicians on treatment \cite{perez2023deep}. Thus, there is an urgent need to improve the understanding of the relationship between clinical and demographic variables with ICH imaging features, in order to gain insights into the underlying factors in prognosis through imaging. Deep convolutional neural networks (CNNs) are able to extract meaningful feature representations from medical images for classification tasks \cite{wang2021review}. Recently, multimodal fusion models are gaining importance to exploit information across different modalities, and to build more precise and robust models \cite{huang2020fusion}. Identifying medical knowledge relevant to image analysis tasks might be complex. While some of this knowledge can be directly learned from the training data, other aspects are not easily captured by the deep learning (DL) model, making it necessary to promote their learning \cite{xie2021survey}.
Transforming medical knowledge into valuable representations to enhance the performance of DL image models also requires a careful understanding of the data \cite{xie2021survey}. 

The focus of this work is on improving feature representation from computed tomography (CT) ICH scans by learning clinical and demographic variables highly correlated with prognosis. The potential of clinical context for ICH prognosis has already been studied in previous works \cite{perez2023deep, shan2023gcs}. Perez \textit{et al.} \cite{perez2023deep} proposed a hybrid model following a joint fusion approach to classify patients into good and poor prognosis, using both CT images and clinical variables. Shan \textit{et al.} \cite{shan2023gcs} simplified the task introducing a multimodal DL algorithm integrating both brain CT image data and Glasgow Coma Scale (GCS) score to improve prognosis. GCS is a clinical variable that describes the extent of impaired consciousness in all types of acute medical and trauma patients, ranging from 3 (worst) to 15 (highest) \cite{jain2018glasgow}. In a related study, Ma \textit{et al.} \cite{ma2023treatment} proposed a generative prognostic model for predicting ICH treatment outcomes utilizing imaging and tabular data. They used a variational autoencoder model to generate a low-dimensional prognostic score, and combined the multi-modality distributions into a joint distribution. All these methods require both tabular and imaging modalities as model inputs.

Current ICH fusion models fail to explore the entanglement between clinical information and medical images, overlooking the extent to which images alone can contribute to prognosis. In routine practice, neuroradiologists (NRs) integrate tabular clinical variables with imaging data to make prognostic assessments and stratify patients \cite{gregorio2019assessment}. Hence, the exploration of how clinical information can be inferred directly from images to enhance prognostic accuracy represents a significant area of interest, particularly in scenarios where certain medical variables cannot be measured due to the patient's condition, such as in intubated ICH patients. Zhou \textit{et al.} \cite{zhou2023novel} showcased the value of incorporating domain knowledge in dermatology proposing a multi-task model to mimic dermatologists' diagnostic procedures, achieving state-of-the-art recognition performance.

In this paper, we propose learning clinical information in the form of discrete binary and ordinal variables to improve feature representation of ICH CT scans in an end-to-end multi-task prognosis model. Our contributions can be summarized as (1) evaluating the clinical and demographic variables with the highest impact on ICH prognosis through machine learning (ML) tabular models, and their best encoding for the multi-task models; (2) introducing the two primary tabular variables driving the prognosis (GCS and age) in two multi-task prognostic image models (binary and ordinal); (3) performing ablations to show the predictive power of the proposed multi-task models; (4) assessing interpretability saliency maps \cite{borys2023explainable} and their alignment with neuroradiologist's knowledge, ultimately comparing the prognostic capabilities of the models with four board-certified NRs.

\section{Materials and Methods}

\subsection{Data}

We used a publicly available single center dataset \cite{10261_275792} of 261 brain CT scans in NIfTI  format, demographic and clinical variables (in the form of tabular data) from the patients' medical history for ICH prognosis, as described in Perez \textit{et al.} \cite{perez2023deep}. Ground truth classification labels are based on hospital survival: 99 patients with Good prognosis (label 0), and 162 patients with Poor prognosis (label 1). 

\subsection{Method}

Previous work conducted on this dataset \cite{perez2023deep} proposed a fusion model concatenating image features extracted by a custom 3D CNN with tabular data extracted by a Dense Neural Network (DNN). However, we identified significant limitations: training curves were close to random guess in the image model, and oversampling was performed in training, validation and test sets. These shortcomings undermined our confidence in the reported results. Thus, we repeated the baseline experiments in Perez \textit{et al.} \cite{perez2023deep} performing 10 fold cross-validation (CV), and limited oversampling to the training set for class balance in prognosis. The 27 CT scans from one test fold were further labelled by four board-certified NRs with an average experience of 7 years (5, 5, 8 and 10 years) for benchmarking. 

\begin{figure}
\includegraphics[width=\textwidth, trim = 0cm 24.2cm 0cm 0.8cm, clip]{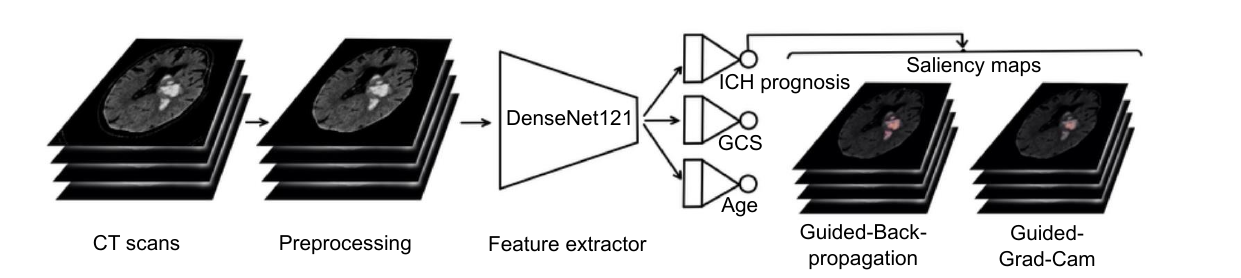}
\caption{Proposed multi-task image model integrating GCS and age as outputs to regularize the learning and enhance the prognosis task. In the saliency maps, brighter colors mean higher importance.}\label{fig:method}
\end{figure}

The proposed method aims to enhance the image model feature representation by learning a shared loss regularization across the main decision-driving variables in the ICH prognosis tabular models. To this end, we first evaluated the prognostic capability of the tabular variables available. Subsequently, we used a 3D DenseNet121 model \cite{monai_consortium_2023_8436376} as feature extractor, and we designed two multi-task image models that aggregated the loss in the prognosis task with the loss of one clinical and one demographic variable, which was back-propagated through the image model. The method is presented in Fig. \ref{fig:method}, and explained below.

\subsubsection{Preprocessing:}
Following \cite{perez2023deep}, numerical variables were normalized using min-max normalization.
In the CT scans, skull was stripped  \cite{hoopes2022synthstrip}, and preprocessing \cite{monai_consortium_2023_8436376} reproduced the steps in Perez \textit{et al.} \cite{perez2023deep} for the best windowing selection. CT scans were downsampled to $301 \times 301 \times 40$ before feeding them to the models.

\subsubsection{Tabular models:} 
First, Perez \textit{et al.} \cite{perez2023deep} DNN tabular model was reproduced, then each variable's importance was evaluated using SHAP values \cite{NIPS2017_7062} (Supplementary Fig. 1). Subsequently, ML tabular models, including Random Forest (RF), XGBoost (XGB), Logistic Regression (LR), and Decision Tree Classifier (DTC), were trained on the most relevant variables to identify the prognostic decision boundaries, which were integrated into the multi-task image models.

\subsubsection{Image models:}
For evaluation purposes, we replicated the baseline image model as described by Perez \textit{et al.} \cite{perez2023deep}. Yet, we also introduced a new baseline model based on a 3D DenseNet121 backbone feature extractor \cite{monai_consortium_2023_8436376}, which mitigates the vanishing-gradient problem and promotes feature reuse \cite{li2021efficient}.
The two key prognostic variables in the tabular models, GCS (clinical) and age (demographic), were incorporated as outputs into the baseline model, transforming it into a multi-task model (Fig. \ref{fig:method}). The aim was to regularize the model's learning process, exploring the shared knowledge between imaging-based prognosis, GCS, and age.

The learning of GCS and age was promoted by encoding them as discrete binary and ordinal variables since, as demonstrated by the DTC, these variables have clear decision boundaries that enable their discretization, which at the same time prevents unnecessary complexity in the multi-task model.
For the binary classification scenario, we followed the DTC's decision boundaries, i.e., GCS from 3 to 8 was set to 1, while GCS from 9 to 15 was set to 0. For the three ordinal classes scenario, we followed Jain and Iverso \cite{jain2018glasgow} GCS division for a common classification of acute traumatic brain injury. Thus, severe GCS from 3 to 8 was set to 2; moderate GCS from 9 to 12 was set to 1; and mild GCS from 13 to 15 was set to 0. Then, we encoded GCS to preserve ordinality ($0 \rightarrow [0,0]$, $1 \rightarrow [1,0]$, and $2 \rightarrow [1,1]$).
Age was binarized according to the decision boundaries in the DTC: age below 80 was established as 0, otherwise it was set to 1.

The first multi-task model predicted prognosis, binary GCS and binary age, hereafter referred to as MT (bin GCS, bin age). The second multi-task model integrated prognosis, three class ordinal GCS and binary age, hereafter referred to as MT (ord GCS, bin age). Both models used a DenseNet121 backbone for feature extraction, and the loss was combined following Eq. \ref{eq:loss} to enhance the feature representation for each task:
\begin{equation}
    \mathcal{L} = \lambda_{prog} \cdot \mathcal{L}_{prog} + \lambda_{GCS} \cdot \mathcal{L}_{GCS} + \lambda_{age} \cdot \mathcal{L}_{age}
    \label{eq:loss}
\end{equation}
where $\mathcal{L}_{prog}$ is the loss in prognosis, $\mathcal{L}_{GCS}$ is the loss in GCS, and $\mathcal{L}_{age}$ is the loss in age. They are all binary cross-entropy losses, since we previously encoded three class ordinal GCS. The hyperparameters in Eq. \ref{eq:loss} were empirically optimized through several experiments: $\lambda_{prog} = 0.4$, $\lambda_{GCS}= 0.3$, and  $\lambda_{age} = 0.3$.

All image models used early stopping with a patience of 20 epochs, evaluated on Balanced Accuracy in validation, and a dropout of $0.2$ in the DenseNet121 feature extractor to prevent overfitting. Small data augmentations applied to the training set included rotations (up to $5\degree$), zoom (up to $10\%$), and Gaussian noise (mean: $0.0$, standard deviation: $0.01$), with a probability of $0.5$, using MONAI \cite{monai_consortium_2023_8436376}. Models were coded in Pytorch (version 1.13.1) \cite{Paszke_PyTorch_An_Imperative_2019} and trained on NVIDIA Tesla T4\textsuperscript{TM} 16GB GPU, utilizing a batch size of 8, three steps of gradient accumulation, and AdamW optimizer \cite{loshchilov2017decoupled} (learning rate of $0.001$, weight decay of $0.0001$).

\subsection{Evaluation}

The performance of the models was assessed on the CV test sets, for a conservative threshold of $0.5$, utilizing the following classification metrics: Area Under the Curve (AUC), Accuracy (Acc.), Balanced Accuracy (B. Acc.), Specificity (Spec.), Negative Predictive Value (NPV), Precision (Prec.), Recall, F1-score (F1-sc.). In the case of the three class ordinal GCS, we used: Acc., B. Acc., Mean Absolute Error (MAE), Root Mean Squared Error (RMSE), Uniform Ordinal Classification index ($A_{UOC}$) \cite{silva2018uniform}, and Cohen Kappa score (quadratic weighted). An ablation analysis was performed to evaluate the contribution of the loss terms in Eq. \ref{eq:loss}.

The impact of the multi-task regularization was further assessed computing interpretability saliency maps, specifically Guided-Backpropagation and Guided-Grad-Cam \cite{gotkowski2021m3d}. To retrieve the most relevant content, saliency maps were normalized and thresholded. 
One of the NRs examined the saliency maps for the baseline and multi-task image models in the same test fold labelled by four NRs.

\section{Results}

\subsection{Tabular models}

Table \ref{tab:tabular} presents the performance of tabular models with mean and standard deviation (SD). XGB was omitted since performance was slightly inferior to RF. SHAP values indicated that GCS and age were the main predictors. LR trained only on GCS and age features reproduced the DNN model by Perez \textit{et al.} \cite{perez2023deep}. Hence, GCS and age were integrated in the proposed multi-task image models.

\begin{table}[]
\caption{Tabular models 10-fold CV performance (mean and SD). Best results are highlighted in \textbf{bold}. Abbreviations: DNN: Dense Neural Network, LR: Logistic Regressor, DTC: Decision Tree Classifier, RF: Random Forest.}\label{tab:tabular}
\begin{center}
\begin{tabular}{|l|l|l|l|l|l|l|}
\hline
\textbf{Model} & \textbf{DNN} & \textbf{RF} & \textbf{LR} & \textbf{LR} & \textbf{DTC} & \textbf{RF} \\ \hline
\textbf{Features} & All & All & GCS & GCS \& age & GCS \& age & GCS \& age \\ \hline
\textbf{AUC}               & $0.79 \pm 0.06$ & $0.78 \pm 0.07$ & $0.70 \pm 0.09$ & $\bm{0.80 \pm 0.07}$ & $0.74 \pm 0.06$ & $0.79 \pm 0.05$ \\ \hline
\textbf{Acc.}          & $0.72 \pm 0.05$ & $0.72 \pm 0.09$ & $0.62 \pm 0.13$ & $\bm{0.73 \pm 0.05}$ & $0.71 \pm 0.06$ & $\bm{0.73 \pm 0.06}$ \\ \hline
\textbf{B. Acc.} & $0.70 \pm 0.04$ & $0.71 \pm 0.09$ & $0.65 \pm 0.10$ & $\bm{0.74 \pm 0.04}$ & $0.69 \pm 0.06$ & $0.72 \pm 0.07$ \\ \hline
\textbf{Spec.}       & $0.66 \pm 0.12$ & $0.64 \pm 0.12$ & $0.77 \pm 0.15$ & $\bm{0.79 \pm 0.10}$ & $0.60 \pm 0.11$ & $0.71 \pm 0.12$ \\ \hline
\textbf{NPV}               & $0.65 \pm 0.12$ & $\bm{0.66 \pm 0.13}$ & $0.53 \pm 0.13$ & $0.63 \pm 0.08$ & $0.63 \pm 0.12$ & $0.63 \pm 0.10$ \\ \hline
\textbf{Prec.}         & $0.79 \pm 0.05$ & $0.78 \pm 0.07$ & $0.79 \pm 0.12$ & $\bm{0.85 \pm 0.06}$ & $0.76 \pm 0.04$ & $0.81 \pm 0.06$ \\ \hline
\textbf{Recall}            & $0.75 \pm 0.14$ & $\bm{0.78 \pm 0.13}$ & $0.5 \pm 0.2$   & $0.70 \pm 0.11$ & $0.77 \pm 0.11$ & $0.74 \pm 0.09$ \\ \hline
\textbf{F1-sc.}          & $0.76 \pm 0.06$ & $\bm{0.77 \pm 0.09}$ & $0.6 \pm 0.2$   & $0.76 \pm 0.06$ & $0.76 \pm 0.06$ & $\bm{0.77 \pm 0.06}$ \\ \hline
\end{tabular}
\end{center}
\end{table}

\subsection{Image models}

\begin{figure}
\centering
\includegraphics[width=0.86\textwidth]{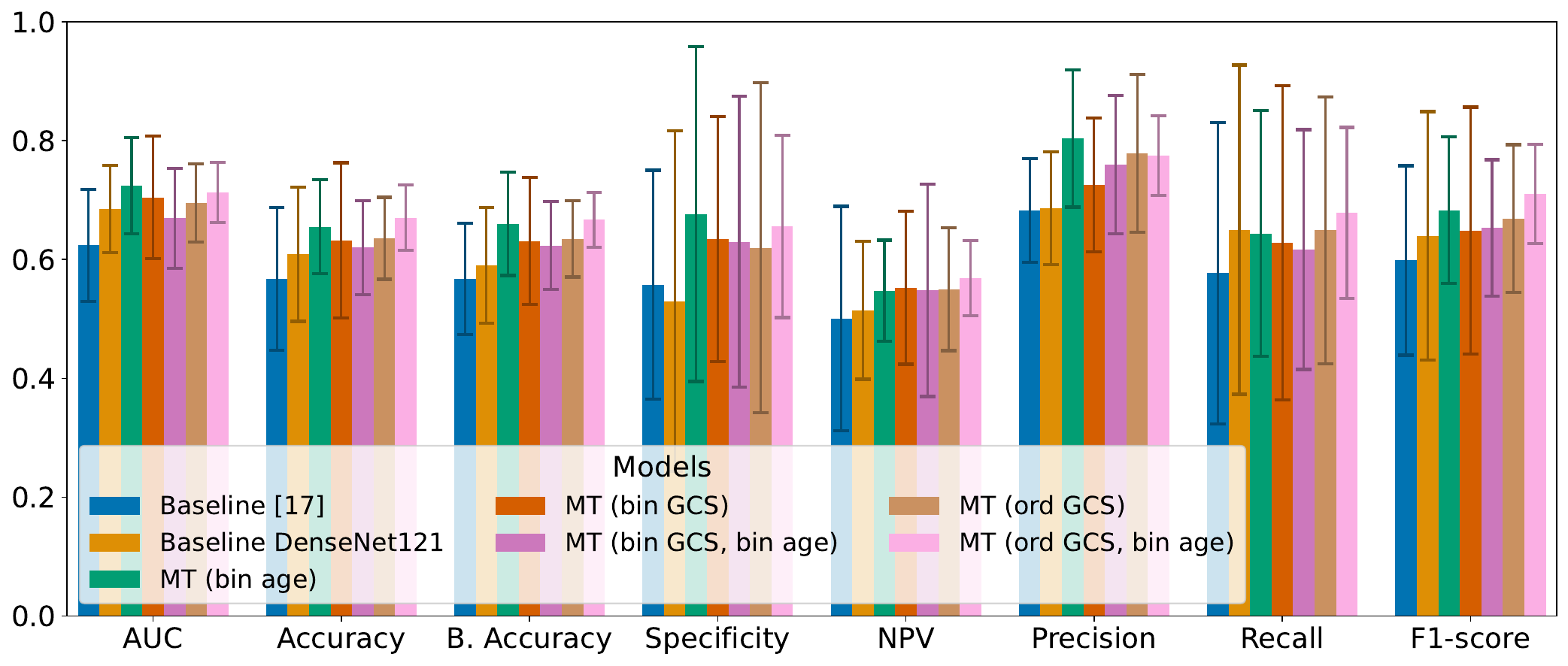}
\caption{Ablation analysis of each component of the loss in our method.}\label{fig:ablation}
\end{figure}

A comparison of the image models' performance with the four NRs in the specific test fold they evaluated is shown in Table \ref{tab:modelVsRadiologists}. This test fold was randomly selected. Ablation analysis on the 10-fold CV test sets is shown in Fig. \ref{fig:ablation}. Separate baseline image models were trained on each outcome variable: prognosis, GCS (binary and ordinal), and age. Results are detailed in Supplementary (Tables 1 and 2).
\begin{table}[H]
\caption{Comparison between models' performance and neuroradiologists (NRs). For the models, $95\%$ confidence intervals estimated by boostrapping are shown in brackets, calculated generating 1000 bootstrap samples from the original test set. For the NRs, mean and SD are given. Best results are highlighted in \textbf{bold}. 
}\label{tab:modelVsRadiologists}
\begin{tabular}{|l|l|l|l|l|l|}
\hline
\textbf{Fold 1} &
  \textbf{Baseline \cite{perez2023deep}} &
  \textbf{\begin{tabular}[c]{@{}l@{}}Baseline\\ DenseNet121\end{tabular}} &
  \textbf{\begin{tabular}[c]{@{}l@{}}MT (bin \\  GCS, bin age)\end{tabular}} &
  \textbf{\begin{tabular}[c]{@{}l@{}}MT (ord \\  GCS, bin age)\end{tabular}} &
  \textbf{NRs} \\ \hline
\textbf{Acc.}          & $0.48 (0.38$-$0.57)$ & $0.63 (0.53$-$0.73)$ & $\bm{0.70} (0.61$-$0.79)$ & $0.66 (0.57$-$0.75)$ & $0.60 \pm 0.06$ \\ \hline
\textbf{B. Acc.} & $0.57 (0.50$-$0.63)$ & $0.52 (0.47$-$0.58)$ & $0.60 (0.54$-$0.67)$ & $\bm{0.67} (0.58$-$0.76)$ & $0.64 \pm 0.05$ \\ \hline
\textbf{Spec.}       & $\bm{0.90} (0.80$-$0.98)$ & $0.10 (0.02$-$0.21)$ & $0.20 (0.08$-$0.33)$ & $0.70 (0.53$-$0.84)$ & $0.80 \pm 0.08$ \\ \hline
\textbf{NPV}               & $0.41 (0.30$-$0.52)$ & $0.50 (0.11$-$0.88)$ & $\bm{1.00} (1.00$-$1.00)$ & $0.54 (0.40$-$0.67)$ & $0.48 \pm 0.05$ \\ \hline
\textbf{Prec.}         & $0.79 (0.60$-$0.95)$ & $0.64 (0.54$-$0.74)$ & $0.68 (0.58$-$0.77)$ & $0.79 (0.67$-$0.89)$ & $\bm{0.81} \pm 0.06$ \\ \hline
\textbf{Recall}            & $0.23 (0.13$-$0.34)$ & $0.94 (0.88$-$0.99)$ & $\bm{1.00} (1.00$-$1.00)$ & $0.64 (0.53$-$0.75)$ & $0.49 \pm 0.12$ \\ \hline
\textbf{F1-sc.}          & $0.36 (0.22$-$0.49)$ & $0.76 (0.68$-$0.83)$ & $\bm{0.81} (0.73$-$0.87)$ & $0.71 (0.62$-$0.79)$ & $0.60 \pm 0.10$ \\ \hline
\end{tabular}
\end{table}

We evaluated the impact of GCS and age on the regularization and explainability of the MT image models computing Guided-Backpropagation and Guided-Grad-Cam saliency maps. Four examples of clinically relevant saliency maps selected by the neuroradiologist (who reviewed all the slices per patient) are depicted in Fig. \ref{fig:saliency}. Due to space limitations, we focus on Guided-Backpropagation, which provided the most insightful activations. Supplementary Fig. 2 includes Guided-Grad-Cam saliency maps for further comparison. Although Fig. \ref{fig:saliency} A is a challenging case to label focusing only on ICH imaging features (some expansivity is shown), the neuroradiologist believes that the MT models’ ability to incorporate GCS (15) and age (78 years) was key to recognizing it as a good prognosis. Interestingly, in Fig. \ref{fig:saliency} B, the MT models highlight the presence of posterior fossa and intraventricular hemorrhage component compared to the baseline model, that does not detect it. Related to Fig. \ref{fig:saliency} C and D, MT (ord GCS, bin age) model additionally detects the lateral component of the subdural hematoma, and shows activations in the adjacent grooves to the subdural hematoma, that is, in the expansivity component, while MT (bin GCS, bin age) and baseline models only detect the medial component of the subdural hematoma. Overall, the neuroradiologist concluded that the MT (ord GCS, bin age) model exhibited fewer arbitrary activations, and probably detected better intraventricular hemorrhage and expansivity signs. The neuroradiologist's criteria and complete assessment of the saliency maps are included in the Supplementary Material (Table 3, Fig. 3).

\section{Discussion and Conclusion}

To the best of our knowledge, this is the first time that the most relevant variables to ICH prognosis (GCS and age) are integrated into an end-to-end multi-task prognostic model to regularize the shared feature representation from CT scans, outperforming baseline models \cite{perez2023deep}, and four board-certified NRs. 
The decision boundaries of GCS and age, established independently by a DTC and supported by the literature \cite{gregorio2019assessment}, focus the learning of the clinical context by allowing the discretization of these variables.
The ablation analysis in Figure \ref{fig:ablation} highlights that the MT (ord GCS, bin age) model exhibits smaller variances compared to all other models, particularly in specificity, recall and F1-score, showing that our approach increased the robustness of the feature representation. 
The metrics were set for a conservative threshold of $0.5$, but could be optimized to maximize recall at the expense of precision, as typically done in clinical application settings \cite{huang2020multimodal}.

\begin{figure}
\includegraphics[width=\textwidth]{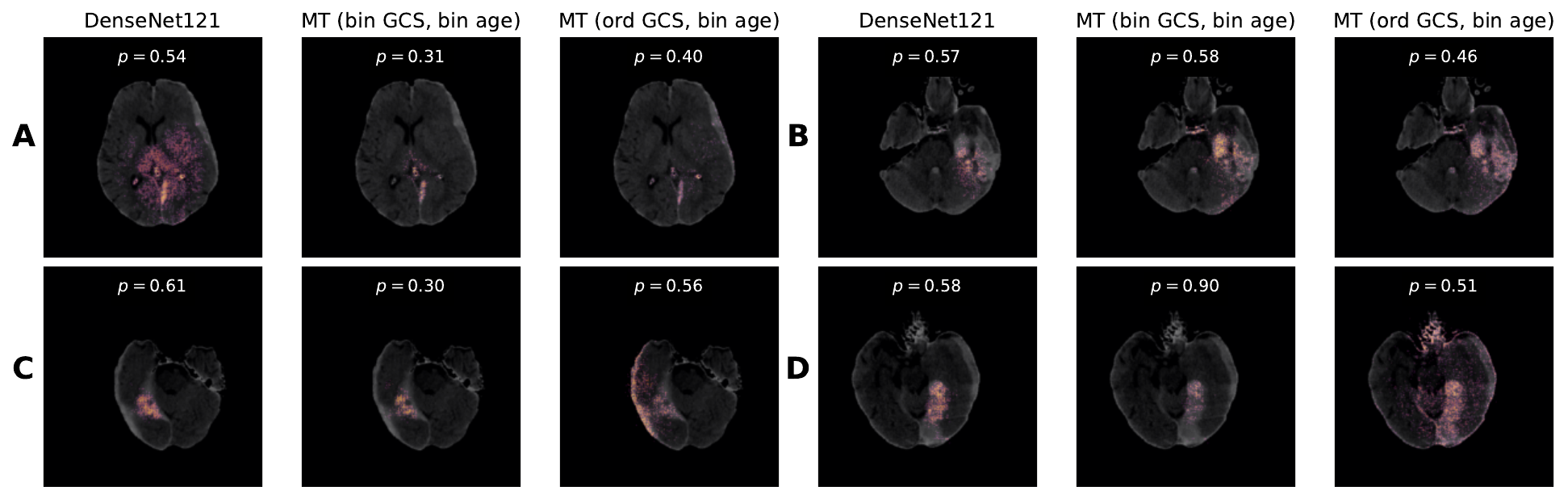}
\caption{Guided-Backpropagation saliency maps ($p$ is the output probability, $p < 0.5$ corresponds to good prognosis). \textbf{A:  Good prognosis.} Correctly labelled by 4/4 neuroradiologists. \textbf{B: Good prognosis.} Incorrectly labelled by 4/4 neuroradiologists. \textbf{C: Poor prognosis.} Incorrectly labelled by 4/4 neuroradiologists. \textbf{D: Poor prognosis.} Correctly labelled by 3/4 neuroradiologists.}\label{fig:saliency}
\end{figure}

The proposed multi-task models demonstrate comparable performance to tabular models, which exhibit strong prognostic capabilities in ICH \cite{gregorio2018prognostic}. However, for reliable prognosis, imaging is essential to evaluate parameters such as the hematoma's volume, expansivity, and possible presence of tumors or other pathologies, all of which contribute more significantly to prognosis and treatment decisions than some other clinical data. Quantifying the information displayed in the image would be more time-consuming (e.g., requiring segmentation of the hematoma for volume quantification), subjective due to the difficulty in precisely characterizing these parameters, and further constrained by the lack of standardization in ICH imaging. 
Direct inclusion of the data in the image itself offers more exhaustive and informative insights.
Thus, our approach mimics clinical decision-making requiring only CT scans as input, and enhances the interpretability of the image model incorporating GCS and age predictions.

The comprehensive evaluation in one of the test folds by four NRs provides additional validation for the multi-task image models, alleviating concerns about the study's single-center data and limited patient sample \cite{la2024don}. The neuroradiologist concluded that the saliency maps across all  models are generally similar. Yet, the MT (ord GCS, bin age) model was more selective or specific, and presented fewer random activations than the MT (bin GCS, bin age) model, and even more so compared to the baseline model. In summary, the baseline model showed more activations, whereas the proposed multi-task models focused more exclusively on the hematoma (or hematomas) and its expansive effect. 
These findings remain qualitative. Thus, future work could focus on quantifying the extent to which the proposed GCS and age variables contribute to image-based prognosis using ground truth segmentations, and assessing model generalizability to new datasets.

In conclusion, we have introduced a novel multi-task method for ICH prognosis leveraging GCS and age, the two main variables driving the decisions in the tabular ICH prognosis models. The proposed multi-task image models regularize the loss using the clinical information embedded in GCS and age outputs, and learn more robust feature representations than state-of-the-art approaches.

\begin{credits}
\subsubsection{\ackname}
M. C. would like to acknowledge the support received by the Ministry of Education of Spain (FPU grant, reference FPU21-04458). The authors would like to acknowledge the support from the project AI4EOSC ‘‘Artificial Intelligence for the European Open Science Cloud” that has received funding from the European Union’s Horizon Europe research and innovation programme under grant agreement number 101058593. The authors acknowledge the neuroradiologists Marta Drake Perez, Elena Marin Diez, and David Castanedo Vazquez from Hospital Universitario Marqués de Valdecilla (Spain) for their contribution evaluating the CT scans.

\subsubsection{\discintname}
The authors have no competing interests to declare that are
relevant to the content of this article. 
\end{credits}
%
%
%
%
\printbibliography

\newpage
\section*{Supplementary material}

\begin{figure}
\includegraphics[width=\textwidth, trim = 0cm 23.8cm 0cm 1.5cm, clip]{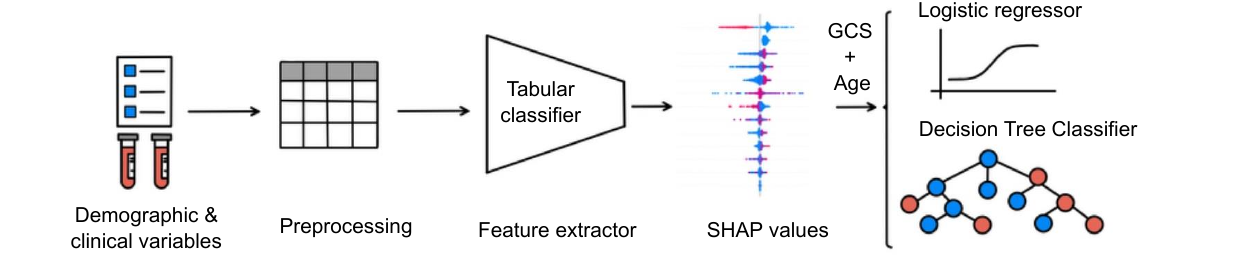}
\caption{Tabular models to identify the most relevant variables driving ICH prognosis predictions, reproduced with logistic regressor and decision tree classifier models trained on the two main variables guiding the decisions: GCS and age.}\label{fig:tabular}
\end{figure}

\begin{table}[]
\caption{Performance in test for the baseline image models trained on each of the outcome variables separately for 10 fold CV (mean and SD).}\label{tab:baselines}
\begin{center}
\begin{tabular}{|l|l|l|l|l|}
\hline
\textbf{Baseline Task} & \textbf{Prognosis \cite{perez2023deep}} & \begin{tabular}[c]{@{}l@{}}\textbf{Prognosis}\\ \textbf{DenseNet121}\end{tabular} & \textbf{Binary GCS}    & \textbf{Binary age}    \\ \hline
\textbf{AUC}      & $0.62 \pm 0.09$      &  $0.69 \pm 0.07$ & $0.71 \pm 0.15$                        & $0.70 \pm 0.13$ \\ \hline
\textbf{Acc.}    & $0.57 \pm 0.12$   & $0.61 \pm 0.11$    & $0.68 \pm 0.08$ & $0.61 \pm 0.15$ \\ \hline
\textbf{B. Acc.} & $0.57 \pm 0.09$ &  $0.59 \pm 0.10$                        & $0.66 \pm 0.13$                        & $0.64 \pm 0.11$ \\ \hline
\textbf{Spec.}   &  $0.56 \pm 0.19$ & $0.5 \pm 0.3$   & $0.71 \pm 0.10$                        & $0.6 \pm 0.2$   \\ \hline
\textbf{NPV}     & $0.50 \pm 0.19$ & $0.51 \pm 0.12$    & $0.86 \pm 0.09$ & $0.86 \pm 0.09$ \\ \hline
\textbf{Prec.}   & $0.68 \pm 0.09$ & $0.69 \pm 0.09$                        &  $0.38 \pm 0.15$ & $0.42 \pm 0.14$ \\ \hline
\textbf{Recall}  & $0.6 \pm 0.3$ & $0.7 \pm 0.3$      & $0.6 \pm 0.2$   & $0.7 \pm 0.3$   \\ \hline
\textbf{F1-sc.}  & $0.60 \pm 0.16$ & $0.6 \pm 0.2$    & $0.45 \pm 0.16$ & $0.50 \pm 0.17$ \\ \hline
\end{tabular}
\end{center}
\end{table}

\begin{table}[]
\caption{Performance in test for the baseline image model trained on three class ordinal GCS model for 10 fold CV.}\label{tab:ThreeClassGCS}
\begin{center}
\begin{tabular}{|l|l|}
\hline
\textbf{Baseline Task}                          & \textbf{Ordinal GCS}   \\ \hline
\textbf{Accuracy}                               & $0.49 \pm 0.11$     \\ \hline
\textbf{Balanced accuracy}                      & $0.44 \pm 0.10$  \\ \hline
\textbf{MAE}                                    & $0.65 \pm 0.13$  \\ \hline
\textbf{RMSE}                                   & $0.98 \pm 0.06$    \\ \hline
\textbf{$\bm{A_{UOC}}$ index} & $0.69 \pm 0.08$ \\ \hline
\textbf{Cohen Kappa score (quadratic weighted)} & $0.19 \pm 0.18$ \\ \hline
\end{tabular}
\end{center}
\end{table}

\begin{figure}
\includegraphics[width=\textwidth]{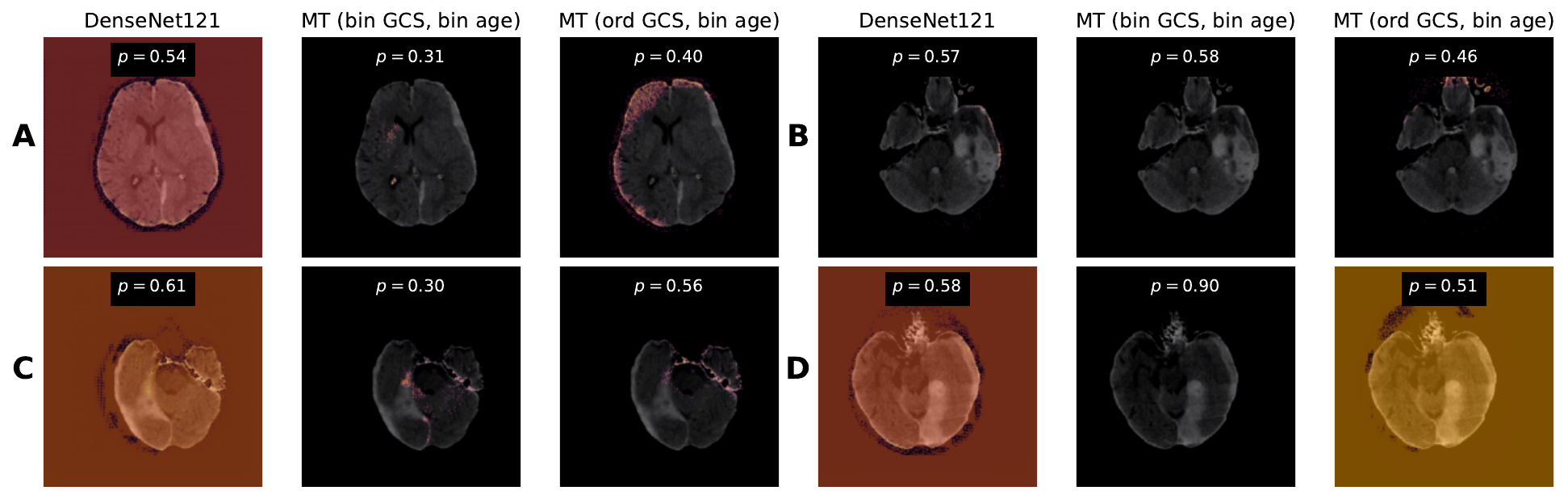}
\caption{Guided-Grad-Cam saliency maps for four example test images ($p$ indicates the output probability of each model). Patient ID is indicated in brackets. \textbf{A:  Good prognosis (34).} Correctly labelled by 4/4 neuroradiologists. \textbf{B: Good prognosis (93).} Incorrectly labelled by 4/4 neuroradiologists. \textbf{C: Poor prognosis (59).} Incorrectly labelled by 4/4 neuroradiologists. \textbf{D: Poor prognosis (140).} Correctly labelled by 3/4 neuroradiologists.}\label{fig:saliency}
\end{figure}

\section*{Neuroradiologist criteria to evaluate the saliency maps}

In relation to the prognosis of intracranial hemorrhage (ICH), Glasgow Coma Scale (GCS), age, hematoma volume, and the presence of intraventricular and/or infratentorial hemorrhage are factors that have been associated with poor prognosis in various studies \cite{hemphill2001score}.

Moreover, in recent years, several potential imaging-specific prognostic factors have been identified. These include the irregularity of hematoma margins and internal hypodensities \cite{serrano2022new}, the spot sign (a hyperdense focus within the hematoma on contrast-enhanced scans that correlates with an increased risk of hematoma expansion) \cite{mata2016dynamic}, and heterogeneous density in subdural hematomas \cite{nakaguchi2001factors}. These factors increasingly underscore the prognostic value of imaging.

The quantity of clinically relevant Guided-Backpropagation saliency maps per patient identified by the neuroradiologist for each model is shown in Table \ref{tab:tab_relevanSM}. Occasionally, no saliency maps were clinically relevant, because there were also activations in the background.
\begin{table}[]
\caption{Number of clinically relevant Guided-Backpropagation saliency maps slices per patient and model according to the evaluation criteria from the neuroradiologist. Patient ID corresponds to the same ID as in the public dataset \cite{10261_275792}.}\label{tab:tab_relevanSM}
\begin{center}
\begin{tabular}{|l|l|l|l|}
        \hline
        \textbf{Patient ID} & \textbf{Baseline} & \textbf{\begin{tabular}[c]{@{}l@{}}MT (bin \\  GCS, bin age)\end{tabular}} & \textbf{\begin{tabular}[c]{@{}l@{}}MT (ord \\  GCS, bin age)\end{tabular}} \\
        \hline
        12 & 9 & 8 & 6 \\ \hline
        20 & 5 & 8 & 12 \\ \hline
        34 & 12 & 14 & 17 \\ \hline
        51 & 13 & 15 & 16 \\ \hline
        56 & 0 & 6 & 6 \\ \hline
        59 & 14 & 7 & 17 \\ \hline
        71 & 16 & 14 & 7 \\ \hline
        90 & 0 & 16 & 9 \\ \hline
        92 & 0 & 11 & 12 \\ \hline
        93 & 14 & 17 & 14 \\ \hline
        106 & 17 & 18 & 17 \\ \hline
        116 & 19 & 19 & 19 \\ \hline
        128 & 18 & 17 & 19 \\ \hline
        137 & 15 & 19 & 12 \\ \hline
        140 & 18 & 16 & 16 \\ \hline
        152 & 7 & 6 & 5 \\ \hline
        170 & 21 & 15 & 21 \\ \hline
        187 & 13 & 14 & 14 \\ \hline
        197 & 0 & 10 & 8 \\ \hline
        203 & 6 & 8 & 6 \\ \hline
        214 & 0 & 6 & 19 \\ \hline
        245 & 15 & 15 & 11 \\ \hline
        251 & 14 & 12 & 12 \\ \hline
        258 & 12 & 0 & 8 \\ \hline
        265 & 12 & 13 & 12 \\ \hline
        268 & 8 & 0 & 6 \\ \hline
        292 & 11 & 13 & 11 \\ \hline
        \textbf{Total} & 289 & 317 & 332 \\ \hline
\end{tabular}
\end{center}
\end{table}

\section*{Comments from the neuroradiologist on the saliency maps}

\begin{figure}
\includegraphics[width=\textwidth]{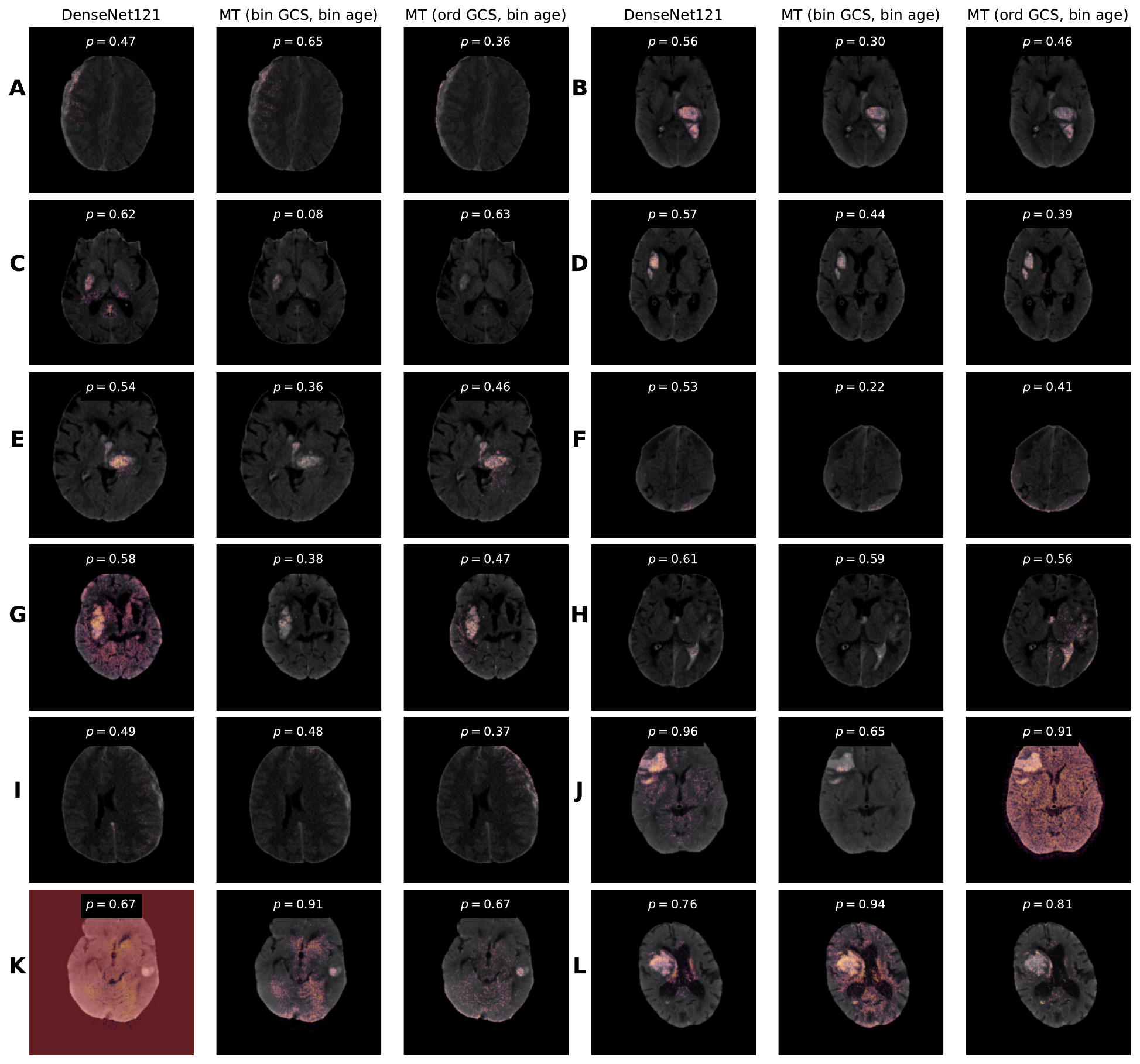}
\caption{Guided-Backpropagation saliency maps ($p$ is the output probability, $p < 0.5$ corresponds to good prognosis). Patient ID is indicated in brackets. \textbf{A:  Good prognosis (51).} Correctly labelled by 4/4 neuroradiologists. \textbf{B: Good prognosis (71).} Correctly labelled by 3/4 neuroradiologists. \textbf{C:  Good prognosis (152).} Correctly labelled by 4/4 neuroradiologists. \textbf{D: Good prognosis (203).} Correctly labelled by 4/4 neuroradiologists. \textbf{E: Poor prognosis (12).} Correctly labelled by 2/4 neuroradiologists. \textbf{F: Poor prognosis (20).} Correctly labelled by 1/4 neuroradiologists. \textbf{G: Poor prognosis (56).} Incorrectly labelled by 4/4 neuroradiologists. \textbf{H: Poor prognosis (106).} Correctly labelled by 4/4 neuroradiologists. \textbf{I: Poor prognosis (128).} Correctly labelled by 4/4 neuroradiologists. \textbf{J: Poor prognosis (170).} Incorrectly labelled by 4/4 neuroradiologists. \textbf{K: Poor prognosis (197).} Correctly labelled by 2/4 neuroradiologists. \textbf{L: Poor prognosis (258).} Correctly labelled by 3/4 neuroradiologists.}\label{fig:saliency_extra}
\end{figure}

The neuroradiologist reviewed all the Guided-Backpropagation saliency maps of the three models: baseline DenseNet121, MT (bin GCS, bin age), MT (ord GCS, bin age), for the 27 patients in the test fold that was further labelled by the four neuroradiologists. Moreover, the neuroradiologist selected an extra set of relevant saliency maps and commented on them. The most representative slice of these saliency maps is depicted in Figure \ref{fig:saliency_extra}. For completeness, we include the neuroradiologist's comments on Fig. \ref{fig:saliency_extra}:
\begin{itemize}
    \item A. MT (ord GCS, bin age) does not show as many activations as baseline model, which is compatible with a good prognosis.
    \item B. All models detect the intraventricular component.
    \item C. Baseline shows more activations, possibly arbitrary, compared to the MT models.
    \item D. Similar activations in all the models. 
    \item E. MT (ord GCS, bin age) detects more the midline intraventricular component compared to the rest.
    \item F. MT (ord GCS, bin age) highlights the highest density component of the bihemispheric subdural hematomas present in the patient. Baseline and MT (bin GCS, bin age) show less useful saliency maps.
    \item G. MT (ord GCS, bin age) highlights hematoma with internal hypodensities. Baseline shows too many activations.
    \item H. MT (ord GCS, bin age) detects all the intraventricular hemorrhage component. Baseline and MT (bin GCS, bin age) do not clearly detect all the intraventricular hemorrhage component or the intraparenchymal hematoma.
    \item I. In addition to the hemorrhage, MT (ord GCS, bin age) also highlighted the expansiveness over the sulci. Baseline did not detect hematoma correctly.
    \item J. In this case the activations in baseline are more meaningful than in the multi-task models.
    \item K. MT (ord GCS, bin age) highlights the expansive effect of a second hematoma located in the posterior fossa (not shown in this slice, but the neuroradiologist spotted it during the revision of all the slices). Too many activations in baseline.
    \item L. MT models highlight more than the baseline model the intraventricular hemorrhage on occipital right horn of the lateral ventricle, but MT (bin GCS, bin age) shows too many activations.
\end{itemize}

\end{document}